
\input phyzzx
\catcode`\@=12 	

\font\eighti=cmmi8                          \skewchar\eighti='177
\font\eightsy=cmsy8                          \skewchar\eightsy='60
\font\eightsl=cmsl8
\font\eightit=cmti8
\def\noblackbox{\overfullrule=0pt}
\noblackbox
\def\bold#1{\setbox0=\hbox{$#1$}%
     \kern-.025em\copy0\kern-\wd0
     \kern.05em\copy0\kern-\wd0
     \kern-.025em\raise.0433em\box0 }
\def\unlock{\catcode`@=11} 
\def\lock{\catcode`@=12} 
\def\Buildrel#1\under#2{\mathrel{\mathop{#2}\limits_{#1}}}
\def\llongrarrow{\hbox to 40pt{\rightarrowfill}}

%
 \newtoks\slashfraction
 \slashfraction={.13}
 \def\slash#1{\setbox0\hbox{$ #1 $}
 \setbox0\hbox to \the\slashfraction\wd0{\hss \box0}/\box0 }
 \unlock
 \def\leftrightarrowfill{$\m@th\mathord-\mkern-6mu%
   \cleaders\hbox{$\mkern-2mu\mathord-\mkern-2mu$}\hfill
   \mkern-6mu\mathord\leftrightarrow$}
 \def\overlrarrow#1{\vbox{\ialign{##\crcr
       \leftrightarrowfill\crcr\noalign{\kern-\p@\nointerlineskip}
       $\hfil\displaystyle{#1}\hfil$\crcr}}}
 \lock
%

{\obeyspaces\global\let =\ }   
%
\def\papersize{	\hsize=35pc\vsize=50pc\hoffset=1cm\voffset=1.3cm
           	\pagebottomfiller=0pc
		\skip\footins=\bigskipamount\normalspace}
\def\lettersize{\hsize=6.5in\vsize=8.5in\hoffset=0cm\voffset=1.6cm
   		\pagebottomfiller=\letterbottomskip
		\skip\footins=\smallskipamount
		\multiply\skip\footins by 3
		\singlespace}
\papers
%
\catcode`\@=11 
\newif\ifletterstyle		
\letterstylefalse 		
\def\letters{\lettersize\letterstyletrue
   \headline=\letterheadline \footline=\letterfootline
   \immediate\openout\labelswrite=\jobname.lab}
\def\iftpub{\afterassignment\iftp@b\toks@}
\def\iftp@b{\edef\n@xt{\Pubnum={UFIFT-HEP--\the\toks@}}\n@xt}
\let\pubnum=\iftpub
\expandafter\ifx\csname eightrm\endcsname\relax
    \let\eightrm=\ninerm \let\eightbf=\ninebf \fi
\catcode`\@=12 
%
   
 

\unlock
\def\eightpoint{\relax
    \textfont0=\eightrm          \scriptfont0=\eightrm
    \scriptscriptfont0=\fiverm
    \def\rm{\fam0 \eightrm \f@ntkey=0 }\relax
    \textfont1=\eighti           \scriptfont1=\eighti
    \scriptscriptfont1=\fivei
    \def\oldstyle{\fam1 \eighti \f@ntkey=1 }\relax
    \textfont2=\eightsy          \scriptfont2=\eightsy
    \scriptscriptfont2=\fivesy
    \textfont3=\tenex          \scriptfont3=\tenex
    \scriptscriptfont3=\tenex
    \def\it{\fam\itfam \eightit \f@ntkey=4 }\textfont\itfam=\eightit
    \def\sl{\fam\slfam \eightsl \f@ntkey=5 }\textfont\slfam=\eightsl
    \def\bf{\fam\bffam \eightbf \f@ntkey=6 }\textfont\bffam=\eightbf
    \scriptfont\bffam=\eightbf     \scriptscriptfont\bffam=\fivebf
    \def\tt{\fam\ttfam \eighttt \f@ntkey=7 }\textfont\ttfam=\eighttt
    \setbox\strutbox=\hbox{\vrule height 4pt depth 3pt width\z@}
    \samef@nt}
\lock
\def\boxit#1{\vbox{\hrule\hbox{\vrule\kern3pt
             \vbox{\kern3pt#1\kern3pt}\kern3pt\vrule}\hrule}}
 \newdimen\str

\def\fboxit#1#2{\vbox{\hrule height #1 \hbox{\vrule width #1
           \kern3pt \vbox{\kern3pt#2\kern3pt}\kern3pt \vrule width #1 }
           \hrule height #1 }}

\def\fillbox#1{\hbox to #1{\vbox to #1{\vfil}\hfil}}
\def\dotbox#1{\hbox to #1{\vbox to 10pt{\vfil}\hss $\cdots$ \hss}}
\def\ggenbox#1#2{\vbox to 10pt{\vss \hbox to #1{\hss #2  \hss} \vss}}


\catcode`\@=11 
\newtoks\foottokens
\let\labelfont=\Tenpoint	
\def\MakeFromBox{\gl@bal\setbox\FromLabelBox=\vbox{\labelfont
     \ialign{##\hfil\cr \the\sendername \the\FromAddress \crcr }}}
\def\smallsize{\relax
\def\eightpoint{\relax
\textfont0=\eightrm  \scriptfont0=\sixrm
\scriptscriptfont0=\fiverm
\def\rm{\fam0 \eightrm \f@ntkey=0}\relax
\textfont1=\eighti  \scriptfont1=\sixi
\scriptscriptfont1=\fivei
\def\oldstyle{\fam1 \eighti \f@ntkey=1}\relax
\textfont2=\eightsy  \scriptfont2=\sixsy
\scriptscriptfont2=\fivesy
\textfont3=\tenex  \scriptfont3=\tenex
\scriptscriptfont3=\tenex
\def\it{\fam\itfam \eightit \f@ntkey=4 }\textfont\itfam=\eightit
\def\sl{\fam\slfam \eightsl \f@ntkey=5 }\textfont\slfam=\eightsl
\def\bf{\fam\bffam \eightbf \f@ntkey=6 }\textfont\bffam=\eightbf
\scriptfont\bffam=\sixbf   \scriptscriptfont\bffam=\sixbf
\def\tt{\fam\ttfam \eighttt \f@ntkey=7 }
\def\caps{\fam\cpfam \tencp \f@ntkey=8 }\textfont\cpfam=\tencp
\setbox\strutbox=\hbox{\vrule height 7.35pt depth 3.02pt width\z@}
\samef@nt}
\normalbaselineskip = 16.60pt plus 0.166pt minus 0.083pt
\normallineskip = 1.25pt plus 0.08pt minus 0.08pt
\normallineskiplimit = 1.25pt
\normaldisplayskip = 16.60pt plus 4.15pt minus 8.3pt
\normaldispshortskip = 4.98pt plus 3.32pt
\normalparskip = 4.98pt plus 1.67pt minus .83pt
\skipregister = 4.15pt plus 1.67pt minus 1.25pt
\def\Eightpoint{\eightpoint \relax
  \ifsingl@\subspaces@t2:5;\else\subspaces@t3:5;\fi
  \ifdoubl@ \multiply\baselineskip by 5
            \divide\baselineskip by 4\fi }
\parindent=16.67pt
\itemsize=25pt
\thinmuskip=2.5mu
\medmuskip=3.33mu plus 1.67mu minus 3.33mu
\thickmuskip=4.17mu plus 4.17mu
\def\thinspace{\kern .13889em }
\def\negthinspace{\kern-.13889em }
\def\enspace{\kern.416667em }
\def\enskip{\hskip.416667em\relax}
\def\quad{\hskip.83333em\relax}
\def\qquad{\hskip1.66667em\relax}
\def\crr{\cropen{8.3333pt}}
\labelwidth=4.5in
\let\labelfont=\Eightpoint
\let\letterhead=\FLOHEAD	
\def\Vfootnote##1{\insert\footins\bgroup
   \interlinepenalty=\interfootnotelinepenalty \floatingpenalty=20000
   \singl@true\doubl@false\Eightpoint
   \splittopskip=\ht\strutbox \boxmaxdepth=\dp\strutbox
   \leftskip=\footindent \rightskip=\z@skip
   \parindent=0.5\footindent \parfillskip=0pt plus 1fil
   \spaceskip=\z@skip \xspaceskip=\z@skip \footnotespecial
   \Textindent{##1}\footstrut\futurelet\next\fo@t}%
\def\attach##1{\step@ver{\strut^{\mkern 1.6667mu ##1} } }
\def\inserttable ##1##2##3%
    {%
    \tbldef {##1}{##3}\goodbreak%
    \midinsert
	\smallskip
	\hbox{\singlespace \hskip 0.5cm
    		\vtop{\parshape=2 0cm 10.8cm 1.3cm 9.5cm
    		      \noindent{\bf\Table{##1}}.\enspace ##3}
		\hfil}
	##2
	\smallskip
    \endinsert
    }
\def\sure{y}
\def\insertfigure ##1##2##3%
    {%
    \figdef {##1}{##3}\goodbreak%
    \midinsert
	\smallskip
	##2
	\hbox{\singlespace\hskip 0.5cm
    		\vtop{\parshape=2 0cm 10.8cm
    			1.6cm 9.2cm \noindent{\bf\Figure{##1}}.
			\enspace ##3}
		\hfil}
	\smallskip
    \endinsert
    }%
\def\references{\par\penalty-300\vskip\chapterskip
        \spacecheck\chapterminspace
	\line{\twelverm\hfil REFERENCES\hfil}
	\nobreak\vskip\headskip\penalty 30000
	\reflist{}}
\def\figures{\par\penalty-300\vskip\chapterskip
        \spacecheck\chapterminspace
	\line{\twelverm\hfil FIGURE CAPTIONS\hfil}
	\nobreak\vskip\headskip\penalty 30000
	\figlist{}}
\def\tables{\par\penalty-300\vskip\chapterskip
        \spacecheck\chapterminspace
	\line{\twelverm\hfil TABLE CAPTIONS\hfil}
	\nobreak\vskip\headskip\penalty 30000
	\tbllist{}}
\def\PH@SR@V{\doubl@true\baselineskip=20.08pt plus .1667pt minus .0833pt
             \parskip = 2.5pt plus 1.6667pt minus .8333pt }
\def\author##1{\vskip\frontpageskip\titlestyle{\tencp ##1}\nobreak}
\def\address##1{\par\kern 4.16667pt\titlestyle{\tenpoint\it ##1}}
\def\andaddress{\par\kern 4.16667pt \centerline{\sl and} \address}
\def\UFL{\address{Department of Physics\break
      University of Florida, Gainesville, FL 32611}}
\def\abstract{\vskip\frontpageskip\centerline{\twelverm ABSTRACT}
              \vskip\headskip }
\def\submit##1{\par\nobreak\vfil\nobreak\medskip
   \centerline{Submitted to \sl ##1}}
\def\doeack{\foot{Work supported by the Department of Energy,
      	contract  DE--FG05--86ER--40272.}}
\def\nsfack{\foot{Work supported by National Science Foundation
      	Grant  PHY 84--16030A01.}}
\def\cases##1{\left\{\,\vcenter{\Tenpoint\m@th
    \ialign{$####\hfil$&\quad####\hfil\crcr##1\crcr}}\right.}
\def\matrix##1{\,\vcenter{\Tenpoint\m@th
    \ialign{\hfil$####$\hfil&&\quad\hfil$####$\hfil\crcr
      \mathstrut\crcr\noalign{\kern-\baselineskip}
     ##1\crcr\mathstrut\crcr\noalign{\kern-\baselineskip}}}\,}
\Tenpoint
}
\newdimen\fullhsize
\newbox\leftcolumn
\def\twoinone{
\smallsize
\def\papersize{
		\voffset=-.23truein
		\vsize=7truein
		\baselineskip=16pt plus 2pt minus 1pt
		\fullhsize=10truein\hsize=4.75truein
                \hoffset=-.54truein
		\skip\footins=\bigskipamount}
\def\lettersize{\voffset=.31truein
	        \vsize=6.38truein
		\baselineskip=16pt plus 2pt minus 1pt
		\fullhsize=10truein\hsize=4.75truein
                \hoffset=-.48truein
		\skip\footins=\smallskipamount
                \multiply\skip\footins by3}
\papers		
\let\lr=L
\output={\if L\lr
		\global\setbox\leftcolumn=\columnbox \advancepageno
		\global\let\lr=R
	 \else  \getitout \advancepageno
		\global\let\lr=L\fi
	 \ifnum\outputpenalty>-20000 \else\dosupereject\fi}
}		
	\def\columnbox{\leftline{
		\vbox{\ifletterstyle\makeheadline\fi
			\pagebody\makefootline}}}
	\def\fullline{\hbox to\fullhsize}
	\def\getitout{\shipout\vbox{\fullline{\box\leftcolumn
		\hfil {\leftline{
		\vbox{\makeheadline
		\pagebody\makefootline}}} }}}
%
\catcode`\@=12 
%
%
%
\newcount	 \ObjClass
\chardef\ClassNum	= 0
\chardef\ClassMisc	= 1
\chardef\ClassEqn	= 2
\chardef\ClassRef	= 3
\chardef\ClassFig	= 4
\chardef\ClassTbl	= 5
\chardef\ClassThm	= 6
\chardef\ClassStyle     = 7
\chardef\ClassDef       = 8
\edef\NumObj	{\ObjClass = \ClassNum   \relax}
\edef\MiscObj	{\ObjClass = \ClassMisc  \relax}
\edef\EqnObj	{\ObjClass = \ClassEqn   \relax}
\edef\RefObj	{\ObjClass = \ClassRef   \relax}
\edef\FigObj	{\ObjClass = \ClassFig   \relax}
\edef\TblObj	{\ObjClass = \ClassTbl   \relax}

\edef\StyleObj  {\ObjClass = \ClassStyle \relax}
\edef\DefObj    {\ObjClass = \ClassDef   \relax}
%
%
\def\gobble	 #1{}%
\def\trimspace   #1 \end{#1}%
\def\ifundefined #1{\expandafter \ifx \csname#1\endcsname \relax}%
\def\trimprefix  #1_#2\end{\expandafter \string \csname #2\endcsname}%
\def\skipspace #1#2#3\end%
    {%
    \def \temp {#2}%
    \ifx \temp\space \skipspace #1#3\end
    \else \gdef #1{#2#3}\fi
    }%
\def\stylename#1{\expandafter\expandafter\expandafter
    \gobble\expandafter\string\the#1}
\ifundefined {protect} \let\protect=\relax \fi
\catcode`\@=11
\let\rel@x=\relax
\def\relaxtest{\rel@x}
\catcode`\@=12
\def\checkchapterlabel%
    {%
        {\protect\if\chapterlabel\relaxtest
	\global\let\chapterlabel=\relax\fi}
    }%
\begingroup
\catcode`\<=1 \catcode`\{=12
\catcode`\>=2 \catcode`\}=12
\xdef\LBrace<{>%
\xdef\RBrace<}>%
\endgroup
%
%
\newcount\equanumber \equanumber=0
\newcount\eqnumber   \eqnumber=0
\newif\ifleftnumbers \leftnumbersfalse

\def\(#1)%
     {%
        \ifnum \equanumber<0 \eqnumber=-\equanumber
	    \advance\eqnumber by -1 \else
            \eqnumber = \equanumber\fi
        \ifmmode\ifinner(\eqnum{#1})\else
        \ifleftnumbers\leqno(\eqnum{#1})\ifdraft{\rm[#1]}\fi
            \else\eqno(\eqnum{#1})\ifdraft{\rm[#1]}\fi\fi\fi
	\else(\eqnum{#1})\fi\ifnum%
	    \equanumber<0 \global\equanumber=-\eqnumber\global\advance
            \equanumber by -1\else\global\equanumber=\eqnumber\fi
     }%
\def\mideq(#1)%
     {%
	\ifleftnumbers \leqinsert{$\(#1)$} \else
	\eqinsert{$\(#1)$} \fi
     }%
\def\eqnum #1%
    {%
    \LookUp Eq_#1 \using\eqnumber\neweqnum
    {\rm\label}%
    }%
\def\neweqnum #1#2%
    {%
    \checkchapterlabel
    {\protect\xdef\eqnoprefix{\ifundefined{chapterlabel}
	\else\chapterlabel.\fi}}
    \ifmmode \xdef #1{\eqnoprefix #1}
        \else\message{Undefined equation \string#1 in non-math mode.}%
	     \xdef #1{\relax}
	     \global\advance \eqnumber by -1
        \fi
    \EqnObj \SaveObject{#1}{#2}
    }%
\everydisplay = {\expandafter \let\csname Eq_\endcsname=\relax
		 \expandafter \let\csname Eq_?\endcsname=\relax}%
%
%
\newcount\tablecount \tablecount=0
\def\Table  #1{Table~\tblnum {#1}}%
\def\tblnum #1{\TblObj \LookUp Tbl_#1 \using\tablecount
	\SaveObject \label\ifdraft [#1]\fi}%
\def\tbldef #1{\TblObj \SaveContents {Tbl_#1}}%
\def\tbllist  {\TblObj \ListObjects}%
%
\def\inserttable #1#2#3%
    {%
    \tbldef {#1}{#3}\goodbreak%
    \midinsert
	\smallskip
	\hbox{\singlespace
	      \vtop{\titlestyle{{\Tenpoint{\caps\Table{#1}}\break #3}}}
    	     }%
	#2
    	\smallskip%
    \endinsert
    }%
\def\topinserttable #1#2#3%
    {%
    \tbldef {#1}{#3}\goodbreak%
    \topinsert
	\smallskip
	\hbox{\singlespace
	      \vtop{\titlestyle{{\Tenpoint{\caps\Table{#1}}\break #3}}}
    	     }%
	#2
    	\smallskip%
    \endinsert
    }%
%
%
\newcount\figurecount \figurecount=0
\def\Figure #1{Figure~\fignum {#1}}%
\def\Fig    #1{Fig.~\fignum {#1}}%
\def\fignum #1{\FigObj \LookUp Fig_#1 \using\figurecount
     \SaveObject \label\ifdraft [#1]\fi}%
\def\figdef #1{\FigObj \SaveContents {Fig_#1}}%
\def\figlist  {\FigObj \ListObjects}%
%
\def\sure{y}
\def\insertfigure #1#2#3%
    {%
    \figdef {#1}{#3}%
    \midinsert
	\bigskip
	\ifx\havefigures\sure
	#2
	\else\fi
	\hbox{	\singlespace
		\hskip 0.4in
		\vtop{\parshape=2 0pt 362pt 32pt 330pt
		      \noindent{\Tenpoint{\caps\Fig{#1}}.\enspace #3}}
		\hfil}
    	\smallskip%
    \endinsert
    }%
\def\topinsertfigure #1#2#3%
    {%
    \figdef {#1}{#3}%
    \topinsert
	\bigskip
	#2
	\hbox{	\singlespace
		\hskip 0.4in
		\vtop{\parshape=2 0pt 362pt 32pt 330pt
		      \noindent{\Tenpoint{\caps\Fig{#1}}.\enspace #3}}
		\hfil}
    	\smallskip%
    \endinsert
    }%
%
%
%
\newcount\theoremcount \theoremcount=0
%
%
%
%
%
%
%
%
%
%
%
%
%
%
\newcount\referencecount \referencecount=0
\newcount\refsequence	\refsequence=0
\newcount\lastrefno	\lastrefno=-1
%
\def\NPrefs{\let\refmark=\NPrefmark \let\refitem=\NPrefitem}

%
\def\refsymbol#1{\refrange#1-\end}%
\def\[#1]#2%
	{%
	\if.#2\rlap.\refmark{\refsymbol{#1}}\let\refendtok=\relax%
	\else\if,#2\rlap,\refmark{\refsymbol{#1}}\let\refendtok=\relax%
    	\else\refmark{\refsymbol{#1}}\let\refendtok=#2\fi\fi%
	\discretionary{}{}{}\refendtok}%
\def\refrange #1-#2\end%
    {%
    \refnums #1,\end
    \def \temp {#2}%
    \ifx \temp\empty \else -\expandafter\refrange \temp\end \fi
    }%
\def\refnums #1,#2\end%
    {%
    \def \temp {#1}%
    \ifx \temp\empty \else \skipspace \temp#1\end\fi
    \ifx \temp\empty
	\ifcase \refsequence
	    \or\or ,\number\lastrefno
	    \else  -\number\lastrefno
	\fi
	\global\lastrefno = -1
	\global\refsequence = 0
    \else
	\RefObj \edef\temp {Ref_\temp\space}%
	\expandafter \LookUp \temp \using\referencecount\SaveObject
	\global\advance \lastrefno by 1
	\edef \temp {\number\lastrefno}%
	\ifx \label\temp
	    \global\advance\refsequence by 1
	\else
	    \global\advance\lastrefno by -1
	    \ifcase \refsequence
		\or ,%
		\or ,\number\lastrefno,%
	    \else   -\number\lastrefno,%
	    \fi
	    \label
	    \global\refsequence = 1
	    \ifx\suffix\empty
		\global\lastrefno = \label
	    \else
		\global\lastrefno = -1
	    \fi
	\fi
	\refnums #2,\end
    \fi
    }%
%
%
%
\def\refnum #1{\RefObj \LookUp Ref_#1
             \using\referencecount\SaveObject \label}%
\def\reflist  {\RefObj \ListObjects}%
%
%
%
%
\newif\ifSaveFile
\newif\ifnotskip
\newwrite\SaveFile
\let\IfSelect=\iftrue
\edef\savefilename {\jobname.aux}%
\def\Def#1#2%
    {%
    \expandafter\gdef\noexpand#1{#2}%
    \DefObj \SaveObject {#2}{\expandafter\gobble\string#1}%
}%
\def\savestate%
    {%
    \ifundefined {chapternumber} \else
	\NumObj \SaveObject {\number\chapternumber}{chapternumber} \fi
        \ifundefined {appendixnumber} \else
	\NumObj \SaveObject {\number\appendixnumber}{appendixnumber} \fi
    \ifundefined {sectionnumber} \else
	\NumObj \SaveObject {\number\sectionnumber}{sectionnumber} \fi
    \ifundefined {pagenumber} \else
	\advance\pagenumber by 1
	\NumObj \SaveObject {\number\pagenumber}{pagenumber}%
	\advance\pagenumber by -1 \fi
    \NumObj \SaveObject {\number\equanumber}{equanumber}%
    \NumObj \SaveObject {\number\tablecount}{tablecount}%
    \NumObj \SaveObject {\number\figurecount}{figurecount}%
    \NumObj \SaveObject {\number\theoremcount}{theoremcount}%
    \NumObj \SaveObject {\number\referencecount}{referencecount}%
    \checkchapterlabel
    \ifundefined {chapterlabel} \else
	{\protect\xdef\chaplabel{\chapterlabel}}
	\MiscObj \SaveObject \chaplabel {chapterlabel} \fi
    \ifundefined {chapterstyle} \else
   	\StyleObj \SaveObject
           {\stylename{\chapterstyle}}{chapterstyle} \fi
    \ifundefined {appendixstyle} \else
	\StyleObj \SaveObject
           {\stylename{\appendixstyle}}{appendixstyle}\fi
}%
\def\Contents #1{\ObjClass=-#1 \SaveContents}%
\def\Define #1#2#3%
    {%
    \ifnum #1=\ClassNum
	\global \csname#2\endcsname = #3 %
    \else \ifnum #1=\ClassStyle
	\global \csname#2\endcsname\expandafter=
	\expandafter{\csname#3\endcsname} %
    \else \ifnum #1=\ClassDef
        \expandafter\gdef\csname#2\endcsname{#3} %
    \else
	\expandafter\xdef \csname#2\endcsname {#3} \fi\fi\fi %
    \ObjClass=#1 \SaveObject {#3}{#2}%
    }%
\def\SaveObject #1#2%
    {%
    \ifSaveFile \else \OpenSaveFile \fi
    \immediate\write\SaveFile
	{%
	\noexpand\IfSelect\noexpand\Define
	{\the\ObjClass}{#2}{#1}\noexpand\fi
	}%
    }%
\def\SaveContents #1%
    {%
    \ifSaveFile \else \OpenSaveFile \fi
    \BreakLine
    \SaveLine {#1}%
    }%
\begingroup
    \catcode`\^^M=\active %
\gdef\BreakLine %
    {%
    \begingroup %
    \catcode`\^^M=\active %
    \newlinechar=`\^^M %
    }%
\gdef\SaveLine #1#2%
    {%
    \toks255={#2}%
    \immediate\write\SaveFile %
	{%
	\noexpand\IfSelect\noexpand\Contents
	{-\the\ObjClass}{#1}\LBrace\the\toks255\RBrace\noexpand\fi%
	}%
    \endgroup %
    }%
\endgroup
\def\ListObjects #1%
    {%
    \ifSaveFile \CloseSaveFile \fi
    \let \IfSelect=\GetContents \ReadFileList #1,\savefilename,\end
    \let \IfSelect=\IfDoObject  \input \savefilename
    \let \IfSelect=\iftrue
    }%
\def\ReadFileList #1,#2\end%
    {%
    \def \temp {#1}%
    \ifx \temp\empty \else \skipspace \temp#1\end \fi
    \ifx \temp\empty \else \input #1 \fi
    \def \temp {#2}%
    \ifx \temp\empty \else \ReadFileList #2\end \fi
    }%
\def\GetContents #1#2#3%
    {%
    \notskipfalse
    \ifnum \ObjClass=-#2
	\expandafter\ifx \csname #3\endcsname
        \relax \else \notskiptrue \fi
    \fi
    \ifnotskip \expandafter \DefContents \csname #3_\endcsname
    }%
\def\DefContents #1#2{\toks255={#2} \xdef #1{\the\toks255}}%
\def\IfDoObject #1#2%
    {%
    \notskipfalse \ifnum \ObjClass=#2
        \notskiptrue\fi \ifnotskip \DoObject
    }%
\def\DoObject #1#2%
    {%
    \ifnum \ObjClass = \ClassTbl	\par\noindent Table~#2.
    \else \ifnum \ObjClass = \ClassFig	\par\noindent Figure~#2.
    \else \ifnum \ObjClass = \ClassRef  \refitem{#2}
    \else \item {#2.}
    \fi\fi\fi
    \ifdraft\edef\temp
       {\trimprefix #1\end}[\expandafter\gobble \temp]~\fi
    \expandafter\ifx \csname #1_\endcsname \relax
	\ifdraft\relax\else\edef\temp {\trimprefix #1\end}%
	[\expandafter\gobble \temp]\fi%
    \else
	\csname #1_\endcsname
    \fi
    }%
\def\OpenSaveFile   {\immediate\openout\SaveFile=\savefilename
               \global\SaveFiletrue}%
\def\CloseSaveFile  {\immediate\closeout
                 \SaveFile \global\SaveFilefalse}%
%
%
\def\LookUp #1 #2\using#3#4%
    {%
    \expandafter \ifx\csname#1\endcsname \relax
        \global\advance #3 by 1
        \expandafter \xdef \csname#1\endcsname {\number #3}%
        \let \newlabelfcn=#4%
          \ifx \newlabelfcn\relax \else
           \expandafter \newlabelfcn \csname#1\endcsname {#1}%
         \fi
    \fi
    \xdef \label  {\csname#1\endcsname}%
    \gdef \suffix {#2}%
    \ifx \suffix\empty \else
        \xdef \suffix {\expandafter\trimspace \suffix\end}%
        \xdef \label  {\label\suffix}%
    \fi
    }%
%
%
%
\newcount\appendixnumber        \appendixnumber=0
\newtoks\appendixstyle	        \appendixstyle={\Alphabetic}
\newif\ifappendixlabel	        \appendixlabelfalse
\def\APPEND#1{\par\penalty-300\vskip\chapterskip
      \spacecheck\chapterminspace
      \global\chapternumber=\number\appendixnumber
      \global\advance\appendixnumber by 1
      \chapterstyle\expandafter=\expandafter{\the\appendixstyle}
\chapterreset
     \titlestyle{Appendix\ifappendixlabel~\chapterlabel\fi.~ {#1}}
      \nobreak\vskip\headskip\penalty 30000}
%

%
%
%
\def\references#1{\par\penalty-300\vskip\chapterskip\spacecheck
       \chapterminspace\line{\fourteenrm\hfil References\hfil}
       \nobreak\vskip\headskip\penalty 30000\reflist{#1}}
\def\figures#1{\par\penalty-300\vskip\chapterskip\spacecheck
     \chapterminspace\line{\fourteenrm\hfil Figure Captions\hfil}
     \nobreak\vskip\headskip\penalty 30000\figlist{#1}}
\def\tables#1{\par\penalty-300\vskip\chapterskip\spacecheck
      \chapterminspace\line{\fourteenrm\hfil Table Captions\hfil}
       \nobreak\vskip\headskip\penalty 30000\tbllist{#1}}
\newif\ifdraft\draftfalse
\newcount\yearltd\yearltd=\year\advance\yearltd by -1900
\def\draft{\drafttrue
    \def\draftdate{preliminary draft:
        \number\month/\number\day/\number\yearltd\ \ \hourmin}%
        \paperheadline={\hfil\draftdate} \headline=\paperheadline
        {\count255=\time\divide\count255 by 60
                   \xdef\hourmin{\number\count255}
        \multiply\count255 by-60\advance\count255 by\time
    \xdef\hourmin{\hourmin:\ifnum\count255<10 0\fi\the\count255} }
	\message{draft mode}  }
%
\def\umdepp{ \Pubnum={$\caps UMD - EPP - \the\pubnum $}
	     \pubtype={$\phantom{blah}$}                 }
\def\slacpub{ \Pubnum={$\caps SLAC - PUB - \the\pubnum $}}
%
%

\def\frac#1/#2{\leavevmode\kern.1em\raise.5ex
		\hbox{\the\scriptfont0
         	#1}\kern-.1em/\kern-.15em
		\lower.25ex\hbox{\the\scriptfont0 #2}}
%
%

\def\frac#1#2{{#1 \over #2}}

\IfSelect \Contents
{-3}{Ref_thooftlargen}{G. 't Hooft,
{\sl Nucl. Phys.} {\bf B72} (1974) 461.}\fi
\IfSelect \Contents
{-3}{Ref_weinbergw}{S. Weinberg and E. Witten,
{\sl Phys. Lett.} {\bf 96B} (1980) 59.}\fi
\IfSelect \Contents
{-3}{Ref_nielsenfishnet}{H. B. Nielsen and P. Olesen,
{\sl Phys. Lett.} {\bf 32B} (1970) 203;
B. Sakita and M. A. Virasoro,
{\sl Phys. Rev. Lett.} {\bf 24} (1970) 1146.}\fi
\IfSelect \Contents
{-3}{Ref_gilest}{R. Giles and C. B. Thorn,
{\sl Phys. Rev.} {\bf D16} (1977) 366.}\fi
\IfSelect \Contents
{-3}{Ref_thornfishnet}{C.B. Thorn,
{\sl Phys. Rev.} {\bf D17} (1978) 1073.}\fi
\IfSelect \Contents
{-3}{Ref_bardakcis}{K. Bardakci and S. Samuel,
{\sl Phys. Rev.} {\bf D16} (1977) 2500.}\fi
\IfSelect \Contents
{-3}{Ref_gilesmt}{R. Giles, L. McLerran, C. B. Thorn,
{\sl Phys. Rev.} {\bf D17} (1978) 2058.}\fi
\IfSelect \Contents
{-3}{Ref_thornfock}{C.B. Thorn,
{\sl Phys. Rev.} {\bf D20} (1979) 1435.}\fi
\IfSelect \Contents
{-3}{Ref_goddardgrt}{P. Goddard, J. Goldstone,
C. Rebbi, and C. B. Thorn, {\sl Nucl. Phys.} {\bf B56} (1973) 109.}\fi
\IfSelect \Contents
{-3}{Ref_kakukikkawa}{M. Kaku and K. Kikkawa,
{\sl Phys. Rev.} {\bf D10} (1974) 1823.}\fi
\IfSelect \Contents
{-3}{Ref_thornsantab}{C. B. Thorn, in
{\it Unified String Theories,} ed. M. Green and D. Gross,
World Scientific Publishing Co. (1986).}\fi
\IfSelect \Contents
{-3}{Ref_thornweeparton}{C.B. Thorn,
{\sl Phys. Rev.} {\bf D19} (1979) 639.}\fi
\IfSelect \Contents
{-3}{Ref_zeldovich}{Ya. B. Zel'dovich,
{\sl Zh. Eksp. Teor. Fiz. Pis'ma Red} {\bf 6} (1967) 883
[{\sl JETP Lett} {\bf 6} (1967) 316].}\fi
\IfSelect \Contents
{-3}{Ref_sakharov}{A. D. Sakharov,
{\sl Dok. Akad. Nauk. SSSR} {\bf 177} (1967) 70
[{\sl Sov. Phys. Dokl.} {\bf 12} (1968) 1040].}\fi
\IfSelect \Contents
{-3}{Ref_adlerinducedgrav}{S. L. Adler,
{\sl Phys. Rev.} {\bf D14} (1976) 379; {\sl Phys. Rev. Lett.}
{\bf 44} (1980) 1567; {\sl Phys. Lett.}
{\bf 95B} (1980) 241; {\sl Rev. Mod. Phys.}
{\bf 54} (1982) 729.}\fi
\IfSelect \Contents
{-3}{Ref_zeeinducedgrav}{A. Zee,
{\sl Phys. Rev. Lett.} {\bf 42} (1979) 417;
{\sl Phys. Rev.} {\bf D23} (1981) 858;
{\sl Phys. Rev. Lett.} {\bf 48} (1982) 295;
{\sl Phys. Lett.} {\bf 109B} (1982) 183.}\fi
\IfSelect \Contents
{-3}{Ref_wittentop}{E. Witten,
{\sl Comm. Math. Phys.} {\bf 117} (1988) 353.}\fi
\IfSelect \Contents
{-3}{Ref_woodardunstablesft}{D. A. Eliezer and R. P. Woodard,
{\sl Nucl. Phys.} {\bf B325} (1989) 389.}\fi
\IfSelect \Contents
{-3}{Ref_kazakov}{V. A. Kazakov,
{\sl Mod. Phys. Lett.} {\bf A4} (1989) 2125.}\fi
\IfSelect \Contents
{-3}{Ref_kazakovm}{V. A. Kazakov and A. A. Migdal,
{\sl Nucl. Phys.} {\bf B311} (1988/89) 171.}\fi
\IfSelect \Contents
{-3}{Ref_brezink}{E. Brezin and V. A. Kazakov,
{\sl Phys. Lett.} {\bf 236B} (1990) 144.}\fi
\IfSelect \Contents
{-3}{Ref_grossm}{D. J. Gross and A. A. Migdal,
{\sl Phys. Rev. Lett.} {\bf 64} (1990) 127;
{\sl Phys. Rev. Lett.} {\bf 64} (1990) 717;
{\sl Nucl. Phys.} {\bf B340} (1990) 333.
{\sl Phys. Rev. Lett.} {\bf 64} (1990) 717.
}\fi
\IfSelect \Contents
{-3}{Ref_douglass}{M. R. Douglas and S. H. Shenker,
{\sl Nucl. Phys.} {\bf B335} (1990) 635.}\fi
\IfSelect \Contents
{-3}{Ref_grossmende}{D. J. Gross and P. F. Mende,
{\sl Phys. Lett.} {\bf 197B} (1987) 129;
{\sl Nucl. Phys.} {\bf B303} (1988) 407.}\fi
\IfSelect \Contents
{-3}{Ref_atickw}{J. J. Atick and E. Witten,
{\sl Nucl. Phys.} {\bf B310} (1988) 291.}\fi
\pubnum={$91-23$}
\date{}
\pubtype={}
\titlepage
\title{Reformulating String Theory with the $1/N$
Expansion\footnote{*}{
Work supported in part by the Department of Energy,
contract DE-FG05-86ER-40272}}
\author{Charles B. Thorn\footnote{\dagger}{Invited
talk at the First International A. D. Sakharov
Conference on Physics, P. N. Lebedev Physical Institute,
Moscow, USSR, May 27-31, 1991.}
}
\address{Department of Physics, University of Florida,
Gainesville, FL, 32611, USA }
\abstract
%
We argue
that string theory should have a formulation for
which stability and causality are evident. Rather than
regard strings as fundamental objects, we suggest they
should be regarded as composite systems of more fundamental
point-like objects. A tentative
scheme for such a reinterpretation
is described along the lines of 't Hooft's $1/N$ expansion and
the light-cone parametrization of the string.
\endpage
\pagenumbers

The discovery of string theory represented a radical
departure from the incremental development of quantum field
theory, which holds that a
physical theory should be formulated
in terms of fields locally defined on space-time.
Since the fundamental entities of string theory are one
dimensional extended objects, important physical features
that are automatic in quantum field theory are not
guaranteed.

\noindent For example, in quantum field theory we can
\singlespace
\point Compute the energy to assess the stability
of the theory, at least in the context of a weak-coupling
(semi-classical) expansion.
\point Easily incorporate Poincar\'e invariance without
disturbing (1).
\point Check that the theory is causal, because the
fundamental entities are point particles (or partons),
and interactions are manifestly local.

\normalspace
In string theory, we have Poincar\'e invariance and
presumably general covariance. But stability and causality
are certainly not manifest, and may be absent altogether.
In perturbation theory, superstring theory is free of
ghosts (one hallmark of acausality) and tachyons (a
hallmark of instability), but the intrinsic nonlocality
of its description obscures the global stability of the
theory. (Expanding $\phi^3$ theory about a locally stable
vacuum reveals no tachyons or ghosts in perturbation theory,
but it is absolutely unstable.)

If stability and causality are present in string theory,
I believe the most natural way to understand this would
be to discover that string should not, at a fundamental
level, be described in terms of one-dimensional objects at
all, but rather as composite structures built from
point-like entities.

\noindent There are many
precedents for extended structures in
quantum field theory:
\singlespace
\pointbegin Vortices in superconductors and superfluids.
\point Nielsen-Olesen vortices in spontaneously broken
abelian gauge theories.
\point Solitons, for example Skyrmions and
\lq t Hooft-Polyakov monopoles.

\normalspace
\noindent In all these cases,
the extended structures exist alongside
ordinary point-like excitations of the fields, whereas
in string theory the extended structures
are everything. Also these features
can't be fit into the framework of a string field loop
expansion.

A more closely parallel local theory is quark confining
QCD. Then all the particle states are those of composite
structures (hadrons). Furthermore, as \lq t Hooft has
shown,\[thooftlargen]\ one
can also establish an expansion of Yang-Mills
theory completely parallel to the string field loop
expansion: the expansion in powers of $1/N_c$ with
$g^2N_c$ fixed. The leading order of this expansion
would give scattering amplitudes displaying an infinite
number of zero width resonances of arbitrarily high
spin, just as the leading order of string theory.
The analogy is so close, in fact, that many authors,
myself included, have wondered whether string theory could
be a $1/N$ expansion of some QCD-like theory.

\noindent But it can't be QCD itself because
\singlespace
\pointbegin It has gravitons. An old folk theorem, proved
relatively recently (1980) by Weinberg and
Witten,\[weinbergw]\ rules out
the existence of massless spin two particles in theories
with a conserved Lorentz covariant energy momentum tensor.
\point It has no trace of point-like (parton) structures.
\point Superstring theory has supersymmetry and is
consistent only in space-time
dimension $D=10>4$.

\normalspace
\noindent Therefore a candidate
underlying theory for superstrings
should
\singlespace
\pointbegin Be generally covariant (probably without
gravitons, since the graviton should be a state of the
string which is to be dynamically generated).
\point Suppress the finite energy-momentum components of the
constituents.
\point Possess supersymmetry and be able to generate
extra dimensions (assuming one starts in four dimensions).
\normalspace

Over a decade ago,
Giles, McLerran, and I gave a fishnet\[nielsenfishnet]\
description of string theory in which
the suppression of finite
momentum constituents was imposed by hand,
and extra dimensions appeared through the
promotion of an internal symmetry to an extra
dimension.\[gilest,thornfishnet,bardakcis,gilesmt,thornfock]\

Our description relied heavily on light-cone
coordinates
$$\eqalign{
x^\pm=&(x^0\pm x^1)/\sqrt{2}\cr
{\bf x}=&(x^2,\cdots,x^{d-1})\cr}
$$
where $d$ is the spatial dimension. To describe dynamics,
$\tau=x^+$ is regarded as the evolution parameter, and
it is also convenient to replace $x^-$ by its conjugate
$P^+\geq0$. A single free particle is then described by
a Schr\"odinger wave function $\psi_{P^+}({\bf x},\tau)$
satisfying
$$i{\partial\over\partial\tau}\psi_{P^+}({\bf x},\tau)
={1\over2P^+}(-{\bf\nabla}^2+m^2)
\psi_{P^+}({\bf x},\tau).
$$

The light-cone string,\[goddardgrt]\ for which
$x^+=\tau$ and the density of the $+$ component
of momentum is the constant rest tension, ${\cal P}^+=T_0$,
 can be described in this
language by discretizing the world-sheet spatial
coordinate $\sigma$ to $M$ lattice sites so that
$P^+=M\epsilon T_0$, after which the states of a
single string can be described by
an $M$ particle wave function
$$\psi_M({\bf x}_1,\cdots,{\bf x}_M,\tau)=
\psi_M({\bf x}_2,\cdots,{\bf x}_M,{\bf x}_1,\tau),$$
governed by the Hamiltonian (note that $P^-$ is
conjugate to $x^+$)
$$P^-={1\over\epsilon}\sum_i^M{1\over2T_0}(-{\bf\nabla_i}^2
+{T_0^2}({\bf x}_{i+1}-{\bf x}_i)^2).
$$
In string field theory one promotes $\psi_M$ to
a dynamical variable through second
quan\-ti\-za\-tion.\[kakukikkawa,thornsantab]\

But instead of second-quantizing $\psi_M$, one
can just as well, and more easily, second-quantize the
constituents. Following 't Hooft's ideas
on the $1/N$ expansion\[thooftlargen]\  we can achieve this by
introducing an $N\times N$ matrix variable
$a_k^{~\ell}({\bf x})$ and its canonical conjugate
${\bar a}_k^{~\ell}({\bf x})=a_\ell^{~k}({\bf x})^\dagger$
so that
$$[a_k^{~\ell}({\bf x}),{\bar a}_m^{~n}({\bf y})]
=\delta_k^n\delta_m^\ell\delta({\bf x}-{\bf y}).
$$
In the large $N$ limit, one can show that the
singlet operators
$${\bar A}({\bf x}_1,\cdots,{\bf x}_M)=
\left({
{1\over N}}\right)^{M/2}
\tr\{{\bar a}({\bf x}_1)\cdots{\bar a}({\bf x}_M)\}
$$
behave as creation operators
for discretized strings.\[thornfock]\
For example, if one considers a Hamiltonian
$$P^-={1\over\epsilon}\int d{\bf x}{1\over 2T_0}
\tr{\bf\nabla}{\bar a}({\bf x})\cdot{\bf\nabla}a({\bf x})
+{1\over\epsilon N}
\int d{\bf x} d{\bf y}{\cal V}({\bf x}-{\bf y})
\tr{\bar a}({\bf x}){\bar a}({\bf y}){a}({\bf y}){a}({\bf x})
\(constitham)
$$
and evaluates its action on a state
$$\ket{\psi}=\left({
{1\over N}}\right)^{M/2}
\tr\{{\bar a}({\bf x}_1)\cdots{\bar a}({\bf x}_M)\}
\psi_M({\bf x}_1,\cdots,{\bf x}_M)\ket{0},
$$
he finds
$$\eqalign{
P^-\ket{\psi}&={1\over\epsilon}
{\textstyle|\sum_i^M(-{\bf\nabla_i}^2/2T_0
+{\cal V}({\bf x}_{i+1}-{\bf x}_i))\psi\rangle}\cr
+&{1\over N\epsilon}\sum_{i=1}^M\sum_{j\neq i,i+1}
{\bar A}({\bf x}_{i+1},\cdots,{\bf x}_{j-1})
{\bar A}({\bf x}_{j},\cdots,{\bf x}_{i})\ket{0}
{\cal V}({\bf x}_i-{\bf x}_j)
\psi_M({\bf x}_1,\cdots,{\bf x}_M).\cr}
$$
The $1/N$ term describes the production of a discretized
string, so $1/N$ should be identified with the
string coupling constant. In the limit that it
vanishes ($N\rightarrow\infty$),
the above equation describes a noninteracting
discretized string provided that the potential ${\cal V}$
is sufficiently attractive to bind. It need not be
a long range harmonic force, because in the continuum
limit $M\rightarrow\infty$ with $P^+=M\epsilon$ fixed, the
finite energy excitations will be
those of a string.\[thornweeparton]\ This is because the
low energy excitations of an $M$ particle discretized string
are $O(1/M)\times$ the two-body level spacing.

If we try to interpret the constituents of string as
particles, we would say that they all carry a fixed
infinitesimal unit of $P^+=\epsilon T_0$.
It is this feature that
in the continuum limit forces the constituents to have
all components of $P^\mu$ infinitesimal. If we compare
$\(constitham)$ to that of an ordinary quantum
field theory, the main difference is that the annihilation
operator would carry an additional label $P^+$,
which could take any positive value. Then quartic terms
involving one creation and three annihilation operators
(and vice versa) would also be allowed. To obtain the
effective hamiltonian $\(constitham)$
from a field theory, the dynamics
would have to suppress this $P^+$ degree of freedom. In
asymptotically free $QCD$ one can compute the constituent
wave-function at short distances and see that there is
no such suppression.

Assuming that one can find a justification for
$\(constitham)$, it is not hard to understand how
internal degrees of freedom can be promoted to
extra ``compactified'' dimensions. This is
illustrated by
a $(\phi^\dagger\phi)^2$ example in which the
constituents are endowed with at abelian charge.\[gilesmt]\
We
showed that summing over all ways charge could flow
through a fishnet diagram produced a 2 dimensional
6-vertex model which has a bosonized description in
terms of a compactified extra dimension.

General
covariance is absent in the above discussion.
But conceivably, the suppression of the finite
energy momentum components can be attributed to
general covariance. For example, if
a generally covariant action does not possess kinetic terms
(such as the Einstein-Hilbert action) for the
metric, it is stationary under variations of the
metric only if the local energy momentum tensor
vanishes,
$$T_{\mu\nu}(x)=0. \(vanishing)$$
Could this be the explanation
for the suppression of finite momentum
components of the string constituents? From the point of view
of string theory, the reason one can work in
light-cone gauge, and in particular can choose the
$P^+$ density constant along the string is, of course,
world-sheet reparametrization invariance. It is
also well-known that world-sheet reparametrization
invariance is intimately linked with general
coordinate invariance, in the sense that the
generators of world-sheet reparametrizations,
the Virasoro operators $L_n$, provide the
Ward-identities associated with target space
general covariance. Perhaps these are hints that
the fundamental theory we seek is a generally covariant
theory in which dynamical
gravitons only appear nonperturbatively.

This means the local theory underlying string
theory should be either a theory with
no curvature terms, as in induced
gravity,\[zeldovich,sakharov,adlerinducedgrav,zeeinducedgrav]\ or
a topological field theory.\[wittentop]\ However, in
the former case, there are (perhaps insurmountable)
technical obstacles to any kind of perturbative
treatment of the theory. It is like the
infinite Newton constant limit of ordinary
gravity coupled to matter. Integrating over the metric
(which is necessary for general covariance)
in the absence of curvature terms imposes, at least
classically, the constraint $\(vanishing)$. But it is
far from clear how such a constraint can be
consistently implemented in a (semi-classical)
loop expansion, especially for space-time dimension
greater than 2. The standard world-sheet BRST formalism
of string theory shows how to handle this constraint
in two space-time dimensions, at least for
conformally invariant theories. This shows that
the constraints are not
necessarily inconsistent. Of course, in this
case the issue of generating gravitons dynamically
does not arise. Unfortunately the techniques that make the two
dimensional case tractable do not carry over to higher
dimensions.

Although we are envisioning a microscopic
formulation of string theory of the above type, it may
not be possible to formulate it properly at the
quantum level without including the dynamical effects
that generate gravitons (\ie\ string formation).\foot{
The $CP(N-1)$ model in two dimensions is somewhat
reminiscent of this situation. There one starts with
an abelian gauge field with no $F^2$ term which would
imply the constraints $J^\mu(x)=0$. But after solving the
theory in the large $N$ limit, an $F^2$ term is generated
and the constraint does not have to be directly dealt with.}
Once gravitons are generated the constraints should
be interpretable as graviton field equations, which
should resolve some of the conceptual problems associated
with them.
Because the graviton in string theory is just a state
of the closed string, it is desirable that the dynamical
mechanism which
generates it also is responsible for string formation.
Since the latter cannot occur in any finite order in
perturbation theory, we should demand the same of the
former.
Divergent but renormalizable
field theories on a curved background
induce an Einstein-Hilbert
term at finite loop order, so in such theories the graviton
would be on a different footing than any strings that might
form.
The best hope for generating string theory
along these lines would therefore
seem to be a generally covariant version of
an ultraviolet finite theory such
as $N=4$ supersymmetric Yang-Mills theory in four space-time
dimensions.

This is all very speculative, but I think the issues
of stability and causality are of such paramount
importance that for string theory to make sense,
it {\it must} have an alternative formulation in
which these properties are not mysterious. I have
tried to suggest a way string theory could come
from quantum field theory because with the latter one can
easily assess these conditions. For example,
the hamiltonian $\(constitham)$ requires ${\cal V}$ to
be attractive, \ie\ negative, for string
formation. With only the exhibited terms the
energy would then be unbounded below and the theory
unstable. One might try to avoid this by making
the constituents fermions, or by adding other
repulsive terms to the energy. Perhaps superstrings
would require such modifications. Probably, string
theory satisfies stability and
causality in a much more subtle way;
conceivably it doesn't satisfy them at all (as
argued by Woodard\[woodardunstablesft]\
in the context of string field
theory).

Finally, let me mention that
in low space-time dimension ($D \leq1$), recent
work\[kazakov,brezink,grossm,douglass]\
has shown how subcritical string theory can arise
from a matrix \lq\lq field theory.'' This is a
prototype, albeit a trivial one, of the reformulation of string
theory I am suggesting.
For $D=1$ the relevant
matrix model is given by a one-dimensional
gauge field interacting
with matter in the adjoint representation.
This suggests looking for generally covariant versions of
such theories in higher dimensions. Again supersymmetric
gauge theories come to mind as they naturally incorporate
extra fields in the adjoint representation.
\medskip
\titlestyle{Postscript, 1994}

Since this preprint was
circulated in late 1991, further insights
and clarifications have developed, which we sketch below.
\pointbegin It should be
emphasized that the system described by
our model hamiltonian (1)
is a collection of {\it nonrelativistic} point
particles (``{\bf String Bits}'') moving in one less spatial
dimension than the strings do. The longitudinal dimension,
corresponding to $x^-$, $P^+$, is nonexistent
for string bits and only appears with string
formation: varying $P^+$ is nothing other than
varying the number of bits in a given string. This manufacture
of an extra noncompact dimension is certainly an intriguing
feature of our scheme.
\point Suppose our string bit model is taken seriously as the
fundamental formulation of string theory. Thermal statistics
of string bits includes a sum over bit number. In the low
temperature limit, when string states should dominate the
thermal partition function, this bit number sum should
become an integral over the $P^+$ value carried
by each string. The measure of this integration
is thus determined unambiguously. Remarkably, it turns
out that this unambiguous measure is exactly the modular
invariant one that
leads to the usual expression for the finite temperature
free energy.
\point If $\epsilon$ is fixed at a very tiny but finite number,
our model of string is very like a polymer with a finite but
very large $O(1/\epsilon)$ ionization energy. At extremely
high temperatures, these polymers should be completely ionized,
giving a high temperature estimate for the free energy
$F\sim-N^2\pi\epsilon/12\beta^{1+(D-2)/2}$, where $D$ is
the space-time dimension perceived by string. Interestingly,
for $D=4$, this matches the (temperature$)^2$ behavior
Atick and Witten\[atickw]\ have argued should characterize the high
temperature phase of string theory. If we require this
matching, any extra (compact) dimensions would have to
arise from an internal degree of freedom carried
by the string bits, along the lines of Ref.[\refnum{gilesmt}].
\point We have observed that the model hamiltonian written in
$\(constitham)$ cannot be stable. A simple way to try to
stabilize it in a way that would retain attractive nearest
neighbor interactions would be to replace the second term with
$${1\over2\epsilon N}\int d{\bf x} d{\bf y}{\cal V}({\bf x}-{\bf y})
\tr:({\bar a}({\bf x}){a}({\bf x})-{a}({\bf x}){\bar a}({\bf x}))
({\bar a}({\bf y}){a}({\bf y})-{a}({\bf y}){\bar a}({\bf y})):$$
where ${\cal V}$ is now taken to be positive (repulsive).
It is easily shown that the nearest neighbor interactions
on a given polymer are attractive for $N>1$, whereas the
interactions between non-nearest neighbors and between bits
on different polymer chains are repulsive. Thus polymer formation
is still favored as long as they are not too dense, and the
hamiltonian is bounded from below.
\point As Gross and Mende\[grossmende]\ and others have emphasized,
string theory should be scale invariant
at short distances.
This suggests that the natural
choice for ${\cal V}$ would
be ${\cal V}({\bf x})=\lambda_0\delta({\bf x})$, which is
scale invariant in 2 transverse dimensions. The attractive
potential $-\lambda_0\delta({\bf x})$ binds with an infinite
binding energy.
However, with a short distance cutoff $1/\Lambda$ on the
delta function
the binding energy $B\sim\Lambda^2 e^{-4\pi/\lambda_0}$
has the required
dependence on $\lambda_0>0$ for dimensional
transmutation. As shown in Ref.[\refnum{thornweeparton}]\ a
random phase approximation
relates this $B$ to the slope $\alpha^\prime$
of the string
Regge trajectories: $\alpha^\prime\approx1/\pi B\sqrt{12}$.
\vskip.3in
\titlestyle{REFERENCES}
\reflist{}
\bye